\documentclass[11pt]{article}
\usepackage{amsmath}
\newcommand{\A}{$\overset{\textrm{\tiny{o}}}{{\textrm{\small{A}}}}$}

\begin{document}

\begin{center}
\Large 
{\bf Hamming distance geometry of a protein conformational space.
 {\large Application to the clustering of a 4 ns molecular dynamics trajectory 
 of the HIV-1 integrase catalytic core.}}

\vskip 1cm
\noindent
Cyril Laboulais$^2$, Mohammed Ouali$^3$, Marc Le Bret$^2$ 
\newline
and Jacques Gabarro-Arpa$^1$ $^2$
\footnote{
Corresponding author. Electronic Address: gabarro@dmi.ens.fr \\
}
\end{center}

\vskip 0.5cm
\begin{flushleft}
\large
\vskip 0.5cm
$^{1)}$ Ecole Normale Sup\'erieure / C.N.R.S.                   \\
\ \ \ Laboratoire Interdisciplinaire de G\'eometrie Appliqu\'ee \\
\ \ \ 45, rue d'Ulm, 75230 Paris cedex (France)                 \\
\vskip 0.5cm
$^{2)}$ LBPA, C.N.R.S. UMR 8532, Ecole Normale Sup\'erieure de Cachan \\
\ \ \ 61, Avenue du Pr\'esident Wilson, 94235 Cachan cedex (France)   \\
\vskip 0.5cm
$^{3)}$ University of Wales, Department of Computer Science                  \\
\ \ \ Group of Computational Biology, Penglais, Aberystwyth SY23 3DB (Wales) \\
\end{flushleft}

\vskip  1cm
{\Large {\bf Abstract}}
\vskip  1cm

Protein structures can be encoded into binary sequences$^{1}$, 
these are used to define a Hamming distance in conformational space: 
the distance between two different molecular conformations is the number 
of different bits in their sequences. 
\par
Each bit in the sequence arises from a partition of conformational space
in two halves. Thus, the information encoded in the binary sequences is also 
used to characterize the regions of conformational space visited by the system.
\par
We apply this distance and their associated geometric structures, to the 
clustering and analysis of conformations sampled during a 4 ns molecular 
dynamics simulation of the HIV-1 integrase catalytic core. 
\par
The cluster analysis of the simulation shows a division of the trajectory
into two segments of 2.6 and 1.4 ns length, which are qualitatively different: 
the data points to the fact that equilibration is only reached at the end 
of the first segment.
\par 
Some length of the paper is devoted to compare the Hamming distance to
the r.m.s. deviation measure. The analysis of the cases studied so far, 
shows that under the same conditions the two measures behave quite 
differently, and that the Hamming distance appears to be more robust than the 
r.m.s. deviation.

\newpage

{\bf Introduction}
\vskip 3mm

Proteins are mesoscopic systems with a complex potential energy surface
that can be characterized by its locally stable conformations, which are 
local potential energy surface minima (PESM). 
There is evidence that proteins move between these conformational substates
and that they can exist in many levels$^{2,3}$. Because of their complexity, 
methods for analyzing these multiple minima can be very useful.
\par
Molecular dynamics simulations (MD) is one of the available techniques for 
simulating the wandering of such systems across their potential energy
surface. To detect the energy substates visited by the resulting trajectory 
we make the assumption that the structures found in PESMs are more 
related to each other than to conformations in other minima;
thus, we follow the approach that consists in using clustering algorithms 
to partition the trajectory into groups of similar conformers, as has been 
done by a number of authors$^{1,4\textrm{-}7}$. This allows to extract useful 
information and at the same time reducing to tractable dimensions the ever 
growing number of data in these computer experiments.
\par
To put related conformations together we need a distance in conformational 
space. In this paper we use a Hamming distance between conformations$^{1}$, 
which is built as follows: 
\begin{itemize}
\item[1)] we assume that there is an order relation among the $N$ atoms 
          in the molecule, 
\item[2)] we form the set of all $ P = \binom{N}{4} $ ordered 4-tuples of 
          atoms, 
\item[3)] for each ordered 4-tuple of atoms \{$a$, $b$, $c$, $d$\}, we 
          calculate the signed volume of the 3-simplex determined by the 
          position of these four atoms:

\vskip 3mm
$ V = 1/6 \; { \begin{vmatrix} 
       x_1^a & x_2^a & x_3^a & 1 \\
       x_1^b & x_2^b & x_3^b & 1 \\
       x_1^c & x_2^c & x_3^c & 1 \\
       x_1^d & x_2^d & x_3^d & 1 
      \end{vmatrix} } \quad  \quad  \quad  \quad $ (1)

\vskip 2mm
\item[4)] from it we define
\vskip 2mm

$ {\chi }(a,b,c,d) = \left\{\begin{array}{rr}
 1 & \mbox{ $V$ $>$ 0}\\
 0 & \mbox{ $V$  =  0}\\
-1 & \mbox{ $V$ $<$ 0}\\
\end{array}\right.  \quad  \quad $ (2)

\end{itemize}

\vskip 3mm

\noindent
the set of all $\chi$ is called the {\bf chirotope}, it is a binary
sequence (if we discard the 0 value as it corresponds to a set of null 
measure). It defines an equivalence relation among conformations: 
two conformations are equivalent if they have the same chirotope. 
When comparing two conformations, whenever a 4-tuple $\chi$ value has 
different signs we call it a {\bf mutation}: the number of mutations 
between two conformations is a Hamming distance.
\par
The Hamming distance and its associated geometrical structures,
as a tool for peering at the energy landscape of a protein, is the subject
of this paper. The Hamming distance has a number of interesting features which 
can be summarized as follows: 
\begin{itemize}
\item[-] It is a true distance obeying the triangle inequality.
\item[-] It is sensitive to chirality.
\item[-] It is computed over a great number of binary topological descriptors.
\item[-] 4-tuples give meaningful geometric objects in the space of 
         conformations, since each one determines a partition of this space 
         into positive and negative halves.
\end{itemize}
\par
The latter can help to put limits on the regions occupied by a particular 
trajectory segment, or to determine the crossing frequency for some boundaries,
as we shall see below.
\par
In what follows we develop our subject around the study of a 4 ns MD 
trajectory of the HIV-1 integrase catalytic core.

\vskip 3mm
{\bf Molecular dynamics}
\vskip 3mm

The simulated molecule is the HIV-1 integrase catalytic core (residues 50-212 
of the integrase); it contains the single mutation F185H and is 
studied in the presence of its cofactor, an Mg$^+$$^+$ ion. It was taken from 
the molecule C of the 1bl3 PDB file$^{8}$ for which the coordinates of all 
the residues in the catalytic loop are available. The end residues (50, 211 
and 212) were added to the structure which was then minimized using our 
quasi-newtonian minimizer$^{9}$.
\par
As the total charge of the protein with the Mg$^{++}$ cation is +1, to ensure 
electric neutrality an extra Cl$^-$ anion was added into the simulation box. 
The 13 nearest water molecules to the Mg$^{++}$ cation were positioned as in 
the PDB file. The EDIT module of AMBER$^{10}$ was modified so that a box of 
given size $50 \times 60 \times 50 $ \A$^3$ could be filled with TIP3P water 
molecules having a density of 1, which amounted to a total of 3822+13 water 
molecules for 2517+2 solute atoms.
\par
The SANDER module of AMBER was used to simulate the dynamics. The force 
field$^{11}$ was complemented with the parameters of Mg$^{++}$ and Cl$^-$ 
ions$^{12}$. Long range coulombic interactions were calculated using the 
particle mesh Ewald method$^{13}$, with a 10 \A { cut} off, and grid size 
$50 \times 60 \times 50 $.
\par
Covalent bonds containing a proton were constrained to their equilibrium 
length using the SHAKE algorithm$^{14}$. The simulation was run in 2 fs steps,
at 300 K constant temperature and 1 atm constant pressure. No correction was 
applied for the neglected long-range van der Waals interactions because they 
are expected to be small.
\par
The protocol to set the water around the protein and ions, and starting the 
simulation was discussed in ref. 15. It consists of the following steps:
\begin{itemize}
\item 5 ps of water heating from 0 K to 300 K with the ions and protein 
      constrained to their initial positions,
\item next the system was restarted at 10 K and heated to 300 K, in 
      intervals of 25 K and 1 ps. As in the previous step only 
      water was allowed to move.
\item After 3.6 ps at 300 K, the system was restarted at 10 K and heated 
      again to 300 K in intervals of 50 K and 3 ps without any constraints. 
\item Finally, after reseting the velocities, we let the system evolve
      in an equilibration step lasting 13 ps at 300 K. 
\end{itemize}
\par
Once this protocol was finished the simulation was started and let run for 
a total simulated time of 4 ns, conformations were sampled every 0.1 ps 
giving a total of 40000 coordinate frames.
The data were visualized and analysed using the program Cadira$^{16}$.

\vskip 3mm
{\bf Analysis of the space of conformations}
\vskip 3mm

   A lot of information can be encoded in the chirotope. In this paper 
we will use it to extract information concerning the trajectory behaviour 
in the regions of conformational space it crosses.
\par
   Each 4-tuple has an associated geometric structure, since for a given 
conformation its $\chi$ value can be either positive or negative, it 
determines a partition of the space of conformations into positive and 
negative halves. 
We define an {\bf event} as the crossing by the trajectory of the boundary 
separating a tuple's positive from negative conformations.
\par
For each tuple we call {\bf major} halfspace the one in which the trajectory 
stays the longest and we call {\bf minor} halfspace the other one.
There is a binary {\bf sequence} associated with each tuple of length the
number of coordinate frames in the trajectory. For each frame the sequence 
has a value 0 or 1 depending if its conformation lies in the tuple's
major or minor halfspace, respectively. An event corresponds to a change 
of value in the sequence.
\par
   We illustrate our point with fig. 1, where conformational space is 
represented, for purposes of illustration, as a two dimensional surface
crossed by lines representing the separation between positive and negative 
halves associated with given tuples.
Suppose that the trajectory evolves between two regions where it can
get momentarily trapped, maybe because they correspond to PESMs, as depicted
in fig. 1: the first one, {\bf R1}, large with shallow boundary walls; 
the second, {\bf R2}, somewhat smaller with a steep boundary. 
\par
   The tuple separation lines within these regions fall into three 
classes: 
\begin{itemize}
\item[-] specific, those that only cross one region, lines beginning 
         by {\bf s} in fig. 1,
\item[-] non-specific that cross more than one region, {\bf ns} lines,
\item[-] and those that fall in between, lines beginning by {\bf b}.
\end{itemize}
See the legend of fig. 1 for typical example sequences associated with 
these tuples.
We say that a given tuple is specific (non-specific) if its associated 
separation boundary is specific (non-specific) respectively.
\par
Among specific tuples we can make a distinction between two extreme cases: 
some that cut a region across the center and others that only scratch the 
border, {\bf sm} and {\bf sw} lines in fig. 1 respectively.
\par
   Concerning the separation lines associated with {\bf sw} tuples we 
see that they are only crossed for a brief lapse of time, the trajectory 
bouncing on the region's walls and inmediately coming back to the interior; 
the sequence of these tuples will only harbor a few events, with a minimum 
of two. A difference between regions {\bf R1} and {\bf R2} is that the latter 
has steeper walls; so, we expect that for {\bf sw} tuples in {\bf R2} the 
average number of frames lying in the tuple minor halfspace will be less 
than for {\bf sw} tuples in {\bf R1} which is shallower.
\par
   On the other hand {\bf sm} tuples are crossed repeatedly while the 
trajectory stays within the region. They give rise to sequences with many 
mutation events within the corresponding sequence subset 
(see legend in fig. 1).
\par
   {\bf b} tuples are crossed when the trajectory goes in between two such 
regions, and as can be deduced from fig. 1, they will give sequences with
mostly one sign for each region.
\par
   The classes of tuples defined above are based on a purely qualitative 
distintion. Actually, there is no a sharp separation between them.
As can be seen from fig. 3 they form a continuum.
\par
   In order to characterize a trajectory interval that stays confined in a
region of the space of conformations, a previous work$^{1}$ introduced a 
parameter: the activity coefficient $D$, which can be calculated with the 
following procedure:
\begin{itemize}
\item[-] 
  first, we calculate the total number of events generated by the specific 
  tuples in this cluster $E_s$
\item[-] 
  second, we calculate the same quantity on every homologous cluster along the
  trajectory and we take the mean value $<E_s>$
\item[-] 
  the activity coefficient $D$ will be the ratio $E_s/<E_s>$.
\end{itemize}
  The assumption is that $D$ will exhibit a value greater than one
for trajectory intervals that remain confined due to repeated bouncing on 
domain walls.
\par

\vskip 3mm
{\bf Clustering}

\vskip 3mm
The basic assumption behind the clustering methodology is that the presence 
of a potential well in conformational space can trap the system. So a
relatively long trajectory segment can evolve in a small region of 
conformational space, giving a set of closely related structures that 
are easily detectable with an appropiate clustering 
algorithm$^{1,4\textrm{-}7}$.
\par
In this paper we employ hierarchical data clustering, which is a process 
that transforms a proximity matrix into a dendogram: a tree representing a 
hierarchy of categories of clusters. There are other approaches to data 
clustering but we employ hierarchichal methods because the energy landscape 
of a protein is supposed to consist of an hierarchical embedding of potential 
wells$^{17\textrm{-}20}$. 
\par
Again we have two categories of hierarchical algorithms: agglomerative and
divisive. Divisive algorithms were tested before$^{1}$ and they were not 
found to perform very well, thus in this paper we restrict ourselves to 
agglomerative ones. These work as follows: for a set of $n$ elements we 
begin with a structure with $n$ clusters, one per object, and grow a 
sequence of clusterings until all $n$ patterns are in a single cluster. 
At each step two clusters are reunited if they optimize a given criteria.
\par
We use the complete-linkage, or farthest-neighbour algorithm,
in which the distance between subsets is the maximum distance between a pair
of elements one in each subset. The complete-linkage hierarchy can be computed
in $O(n\;{\textrm{log}}^2 n)$ time$^{21}$.
One reason for choosing this particular algorithm is that it enhances cluster 
separation and we think it is more adequate for studying long trajectories.
\par
One problem with clustering algorithms is that clusters will always
be found even if the data are entirely random,
so we need procedures that help us to determine which clusters 
do really correspond to PESMs. Here we employ two different methods: 
the first one, as explained in the previous section, the likelyhood
of PESM activity for each cluster is estimated with the parameter $D$.
Following our analysis of conformational space, a PESM should display 
a greater than average number of specific events specially in the range 
of {\bf sw} tuples, thus a $D$ value high above 1 can be considered 
a reliable indicator.
In order to validate our conformational space methods, we used energy 
minimization as our second test procedure, as it is described in the 
following section.
\par
   Two distance matrices were built from the set of C$_\alpha$ carbons
exclusively$^{6,7}$: one based on the Hamming distance as described above, 
and another based on the r.m.s. deviation (RMSD). 
The latter is given by$^{22}$:

$ \textrm{RMSD} = \underset{R}{\textrm{min}}\:(\frac{1}{N}
  \overset{N}{\underset{i}{\sum}}
  \|R.\textrm{x}^a_i-\textrm{x}^b_i\|^2)^{1/2} $

\noindent
where $N$ is the total number of atoms, $\textrm{x}^a_i$ and $\textrm{x}^b_i$
denote the atomic coordinates of the $i$th atom of structures $a$ and $b$ in
a common center of mass coordinate system, the distance is the minimum taken
over all $3 \times 3$ rotation matrices $R$.
\par
The reason for choosing the RMSD from the set of available distance 
measures$^{5}$, is that, as for the Hamming distance, it applies to 
a set of equivalence classes in conformational space: two conformations
{\it {\bf a }} and {\it {\bf b }} are in the same equivalence class 
if there exists a rotation and a translation in three-dimensional space 
that transform {\it {\bf b }} into {\it {\bf a }}.
\par
As defined above the distance between any two such equivalence classes 
is the minimum r.m.s. between any two conformations one in each class.
\par
Moreover, the classes belonging to this set are obviously embedded 
in the equivalence classes defined by the chirotope, as lower dimensional 
manifolds.
\par
Comparison with other similarity measures, like the r.m.s. difference between 
corresponding torsion angles, and the r.m.s. deviation of intramolecular
distances$^{5,6}$, was not considered meaningful since they do not operate 
on cartesian conformational space, like the RMSD and the Hamming distance.

\vskip 3mm
{\bf Energy Minimization}
\vskip 3mm

   A PESM acts as an attractor on the trajectory which can get stuck in it 
for a while. The structures sampled along the corresponding segment of the 
trajectory will form a cluster of similar conformations. The contrary may 
not be true: a cluster of similar conformations does not necessarily 
indicate the presence of an energy surface minima. We use energy 
minimization as an alternative method to see if a cluster represents a 
feature of the energy lanscape. 
\par
   In a first approximation, a PESM can be thought as a paraboloid shaped
structure, that becomes narrower as we approach towards the minima. Thus,
upon moderate minimization the distance between any two conformations in a 
cluster will decrease.
\par
   We performed energy minimization on structures belonging to clusters
of homologous conformers using the module SANDER from AMBER$^{10}$. 
Minimization was perfomed with the ewald option, up to a maximum r.m.s. 
gradient of 0.4, which is somewhat high. However, enhanced minimization 
is not of much use in our case since the effective potential surrounding 
the protein depends upon the organization of the solvent and ions. Continued 
minimization will disrupt this structure by inducing extensive reorganization 
of the water network.
\par
This precludes the use of energy minimization to estimate the size of features 
in the energy lanscape, as well as identifying smaller minima within PESMs.
But it is enough to observe a shrinking of r.m.s. deviations (RMSDS) between
elements of the cluster which is an indicator of the presence of a PESM,
otherwise in this paper we did not attempt at resolving the fine structure.

\vskip 3mm
{\bf Results}
\vskip 3mm

From the 28342440 4-tuples present in our system 25 percent undergo at least 
one mutation during this dynamics run. Between conformations 5 ps away we 
observe a minimum distance of 525922 and a maximum of 2479784.
\par
The dendogram, fig. 2, clearly separates the frames into two big clusters 
we call {\bf A} and {\bf B}: 
cluster {\bf A} harbors the first 2.6 ns, cluster {\bf B} spans 1.4 ns 
until the end of the run (see table I for a description of the main clusters).
Cluster {\bf A} has an activity coefficient of 6.86 and is dominated by 
a big subcluster: {\bf A}$_0$, 1.4 ns in length with a staircase structure. 
Interestingly {\bf A}$_0$ contains the clusters at the deepest level 
in the hierarchy which are organized around the initial structure: 
the cluster {\bf A}$_0$$_0$ spanning the first 45 ps has an activity 
coefficient of 11.07, it is embedded in cluster {\bf A}$_0$$_1$ which spans 
the first 255 ps with a coefficient 7.76 (see fig. 3a for a plot showing 
the spectrum of this cluster). This shows that the region around the initial 
structure exhibits PESM activity and is well separated in conformational space 
from the rest of the trajectory.
\par
As we climb cluster {\bf A}$_0$ up to the top the activity coefficient 
gradually disminishes up to a minimum of 1.94 at the top, where the
{\bf sw}-tuple activity has completely disappeared from the spectrum,
but there still remains a high distribution of {\bf sm}-tuples (see fig. 3b). 
This trend continues past {\bf A}$_0$ up to {\bf A} which also displays 
little {\bf sw}-tuple activity (see fig. 3c).
\par
It is difficult to discern a prominent feature inside {\bf A}$_0$; 
it seems as if the trajectory could not get away from the artificially created
PESM around the energy minimized initial structure, and this PESM structure 
were slowly decaying.
\par
Interspersed within the segment of the trajectory spanned by {\bf A}$_0$ 
there are clusters {\bf A}$_1$, {\bf A}$_3$ and {\bf A}$_5$ which exhibit 
genuine PESM activity, since they have a well defined {\bf s}-tuple 
activity as well as RMSDS upon minimization.
However, analysis of their tuple spectrum (fig. 3d) shows a lack of 
significant {\bf b}-tuple activity for these clusters; this seems to 
indicate that they could arise as a bulge from the main cluster: the 
trajectory gets momentarily out of {\bf A}$_0$ and then plunges back.
\par
Consistent with this observation is the fact that {\bf A}$_1$, {\bf A}$_3$ 
and {\bf A}$_5$ lie higher in the 
cluster hierarchy than {\bf A}$_0$, indicating that they are somewhat 
distant from it in conformation space. 
\par
Clusters which are relatively close to {\bf A}$_0$, {\bf A}$_2$ and 
{\bf A}$_4$ display RMSDS, but like {\bf A}$_0$ their 
{\bf sw}-tuple activity is negligible. This trend goes up to the parent 
cluster {\bf A} whose spectrum is rich in {\bf sm}-tuples and 
{\bf b}-tuples, indicating that it is well separated from the rest of the 
trajectory as well as some internal structure. However its {\bf sw}-tuple 
is no more than average (see fig. 3c). Only the clusters around the initial 
energy minimized structure {\bf A}$_0$$_0$ and {\bf A}$_0$$_1$ and the 
bulge clusters {\bf A}$_1$, {\bf A}$_3$ and {\bf A}$_5$ display any 
{\bf sw}-tuple activity.
\par
This contrasts with the situation in {\bf B} whose spectrum (depicted in 
fig. 3e) shows activity for every class of tuples all along the hierarchy, 
down to the bottom. Besides all clusters in this part of the dendogram, 
from {\bf B}$_1$ to {\bf B}$_5$, appear to be well separated. They represent
continuous trajectory segments, and all, except {\bf B}$_3$, exhibit PESM 
activity either from the tuple spectrum or from RMSDS (see table I).
{\bf B}$_3$ is an exception, it spans a continuous strecth of the trajectory 
from 2.97 to 3.53 ns. No PESM activity of any kind is detected for this 
cluster. Activity is only detected down the hierarchy for small clusters 
spanning less than 100 ps. These data are not shown since at this point 
our dendogram attains its limit resolution. 
\par
As before, the end of the trajectory, spanned by cluster {\bf B}$_5$, 
(see fig. 2 and fig. 3f) appears at the deepest level in cluster {\bf B}.
\par
When we apply tuple spectrum analysis separately for clusters {\bf A} and 
{\bf B} we obtain qualitatively the same results on both, except for cluster 
{\bf A}$_5$. The activity coefficient of this cluster increases considerably, 
but this is not unexpected since it lies at the extremity of the trajectory 
segment of cluster {\bf A}.
From this spectrum invariance of tuples encoding the ocupation structure of
conformational space, for {\bf A} and {\bf B}, we can conclude that there is 
little interference between the two regions and that the two consecutive 
segments of the trajectory they contain are well separated in conformational 
space.
\par 
In the upper right part of fig. 4 we have plotted the Hamming distance matrix,
clusters {\bf A} and {\bf B}  (see table I) are easily identified by simple 
visual inspection.

\vskip 3mm
{\bf Comparison with r.m.s. clustering}
\vskip 3mm

\par
Clustering can be performed using other measures than Hamming distance,
as a comparison we have used the RMSD$^{22}$, which is a widely used 
similarity measure.
\par
The dendogram which results from r.m.s. deviation clustering, using the same
protocol as above, can be seen in fig. 5a. 
It is composed of two big clusters, superficially each one looks like 
{\bf A}$_0$ in fig. 2: they are an aggregation of small clusters of 
average length 100 ps, no other visual features can be discerned. A closer 
look to the data gives the following:
\begin{itemize}
\item The two main clusters correspond to discontinuous trajectory segments.
\item Clusters {\bf A}$_0$$_0$ and {\bf B}$_5$ which contain the extremities
      of the trajectory, appear at branching levels relatively close to the 
      root of the tree, moreover {\bf B}$_5$ comes in two separated clusters.
      Indeed the three is not well balanced.
\item In fig. 5a we have underlined clusters which correspond to fragments of 
      clusters identified in the dendogram of fig. 2 (with the exception of 
      {\bf A}$_0$). Many previously identified clusters (harboring continuous 
      trajectory segments) appear fragmented, some of them scattered over 
      many branchs. 
\end{itemize}
Obviously the protocol used for the Hamming distance gives poor results when
used with the RMSD.
\par
From a previous work using the same protocol$^{1}$, it was reported more 
meaningful results with RMSD clustering. However, they were obtained for 
a 1 ns trajectory, which is four times shorter than the present one, and 
the protein studied was the bovine pancreatic trypsin inhibitor$^{23}$, 
which is about three times smaller than the one studied in the present work.
Also in that occasion the authors deemed not necessary to report that the
dendogram tree was poorly balanced and that the main branches harbored
some discontinuities, two of the features identified above.
The result seemed acceptable because the analysis of the low level clusters,
using discrete geometry tests, closely matched the ones that had been 
previously identified using Hamming data clustering$^{1}$.
Seen in the light of the present results it seems that the work with BPTI 
attained the limit resolution that can be obtained with the above method, 
using the RMSD.
\par 
For sake of comparison, we have plotted in fig. 4 the Hamming vs the r.m.s. 
distance matrices. Although both display the most salient features, 
the Hamming distance matrix appears to be much more detailed at the fine 
grained level. This suggests that better results could be obtained with RMSD 
if resolution were improved.
\par
One way of improving resolution is to calculate the distance matrix using 
more atoms than the C$_\alpha$ carbons, two possibilities were tested: 
one with all the backbone heavy atoms (carbonyl carbon, amide nitrogen and
C$_\alpha$), another with all the protein heavy atoms.
Neither works, with the backbone heavy atoms option the resulting tree had 
a less monotonous appearence and had improved balancing, but again it failed 
to resolve the main clusters, and many smaller clusters 
(between 150 and 500 ps in length) corresponded to discontinuous trajectory 
segments. 
\par
The option involving all heavy atoms gave the worst results, the reason 
is that individual regions along the protein jump between conformational 
substates, and the combination of all these transitions distribute uniformly
through conformational space, thus smearing the data$^{6}$.
\par
It may also happen that the complete-linkage is not the best clustering 
algorithm available $^{7}$, at least when used together with RMSD. 
As in ref. 7 we have tried the average distance criterion,
where the distance between two clusters is taken as the average distance
between the points in one cluster and the points in the other.
\par
The resulting dendogram, see fig. 5b, yields better results for the RMSD
(with all the backbone heavy atoms option):
at least the algorithm correctly identifies the main clusters {\bf A} and 
{\bf B}, so we can compare with the results from previous section.
It also identifies the clusters corresponding to well defined PESMs:
{\bf A}$_0$$_0$, {\bf A}$_0$$_1$, {\bf A}$_1$, {\bf A}$_5$, and every cluster 
on the {\bf B} side. The {\bf B} cluster obtained is quite close to 
the one in fig. 2.
\par
The same cannot be said of the {\bf A} subtree: to begin with it is not 
correctly balanced, {\bf A}$_0$$_0$ and {\bf A}$_0$$_1$ appear close
to the root; also, the two halves of {\bf A}$_2$ and {\bf A}$_3$ from fig. 2,
are in two separate branches; finally, {\bf A}$_4$ is scattered over 
many branches.
We think that this protocol can be improved to give still better results, 
but in the scope of the present work, we do not see the point in getting 
any further.
\par
On the other hand the average distance cluster algorithm with the Hamming
distance gives a result quite similar to the one discussed in the previous 
section.

\vskip 3mm
{\bf Discussion}
\vskip 3mm

After performing conformer structure clustering of an MD trajectory based 
on a Hamming distance, the analysis of the dendogram shows a very 
strong PESM at the begining
of the dynamics. It surrounds the energy minimized structure created
at the start of the run and lies at the bottom of the hierarchy of a 
staircase shaped featureless cluster that spans 1.4 ns of the trajectory.

This PESM dominates the initial part of the simulation,
analysis reveals a slow decay: as we climb the hierarchy of
clusters along the stairs the activity disminishes steadily, this is
due to a waning of the {\bf sw} part of the tuple spectrum, this 
low {\bf sw} activity cluster indicates an absence of sharp boundary walls.
The initial PESM has progressively become shallower and structureless. 

This process is ponctuated by incursions into smaller PESMs with high 
{\bf sw} activity, pointing to the presence of sharp boundaries, 
but totally absent {\bf b} activity (see fig. 3d), indicating that these 
PESMs may be a temporary bulge.
\par
This behaviour may be attributed to the effect of the protocol used in
preparing the dynamics run; the whole system is assembled from separate
subsystems: the crystallographic structure for the protein and a cube 
containing a Monte Carlo generated distribution of water, which are relaxed 
separately. The result is that the final system is only locally 
equilibrated. The use of the Ewald summation method to calculate long
range interactions contributes a great deal, since it enhances the 
sensitivity of the system to global conditions.
The anomalous structure in the protein dendogram first half can be attributed 
to the fact that the system is only locally equilibrated, global equilibrium 
being attained much later in the dynamics run.
\par
This kind of behaviour has already been reported$^{24}$. There is however 
a difference between their results and ours: in the first half of the 
dynamics, their conformations appear to be much more uncorrelated than ours 
(see fig. 3 in ref. 24). 
We may attribute this difference to the fact that, in our case, the Ewald 
summation method is more constraining for the protein structure. Thus 
explaining the attraction exerted by the initial structure during the early 
part of the simulation.
\par
Cluster {\bf B} has a number of well separated subclusters showing clear PESM 
activity. They correspond more to the expected behaviour of a protein 
with the trajectory jumping between metastable substates$^{17,20}$. 
Visual inspection of the distance matrix plot in fig. 4 shows that there is 
a clear separation between clusters {\bf A} and {\bf B}. The same can be said 
after analysing their respective tuple spectrum (figs. 3c and 3e) showing 
a great number of specific {\bf b} tuples, which indicates that they lie in 
well separated regions of conformational space.
\par
The dendograms built using a Hamming distance matrix reflect the degree 
of separation in conformational space between different trajectory segments: 
they are balanced binary trees over the trajectory.
\par
One conclusion that can be drawn from the comparison between Hamming and 
RMSD clustering is that both measures are not equivalent: their behaviour 
can be radically different under a given clustering algorithm.
\par
Another conclusion is that in the scope of the cases tested to this day, 
the Hamming distance measure appears to be more robust than RMSD:
it can perform fairly well with smaller data sets, and gives 
consistent results at least with the two algorithms employed here.
In the present work the RMSD seems to work well only with averaged distances,
something that was also reported in ref. 7.
We see three reasons for that.
\par
First, hierarchical clustering algorithms rely upon iterative procedures, 
which in many cases are known to be extremely sensitive to initial 
conditions $^{25}$, thus the observed instability of the double linkage 
algorithm with the RMSD is not an uncommon behaviour.
\par
Second, the r.m.s. deviation is built on $N$ terms, while Hamming distance 
is evaluated over $ \binom{N}{4} $ binary descriptors, roughly $O(N^4)$. 
The fact that increasing the number of atoms improves the results of RMSD 
calculations seems to support this conclusion. Also, the Hamming distance 
matrix in fig. 4 appears to be more detailed at the fine grained level than 
the r.m.s. deviation matrix.
\par
Third, the hierarchical structure of conformational spaces is built-in within 
the Hamming distance. This follows from its physical interpretation as the 
minimum number of topological transitions (i.e. a point crossing the plane 
defined by three other), that have to be effected to transform 
one conformation into another. This has the effect that no matter how much 
we change this discrete distance by adding extra points, it remains that 
the old structure is always embedded within the new one. 
This feature helps to understand the observed good scaling behaviour.

\vskip 3mm
{\bf Conclusion}
\vskip 3mm

In this paper we aimed at describing a new distance measure for molecular
conformational space, and we used it for clustering the conformations 
sampled in the MD simulation of the HIV-1 integrase catalytic core. 
\par
We have described how molecular conformations can be characterized by 
a binary sequence: the chirotope, in which each bit is the value of a binary 
topological descriptor. The distance between any two conformations is the 
number of different bits between the corresponding sequences; this is the 
definition of a Hamming distance, which is a discrete one.
\par
We have shown that, in the cases studied so far, this measure works 
differently than the r.m.s. deviation, which is a currently employed 
similarity measure, and most of the time works better.
We see two reasons for this:
\begin{itemize}
\item[1)] the much greater number of descriptors used in computing our 
          distance, thus embodying far more information,
\item[2)] the fact that our topological descriptors are related to well 
          defined geometrical structures in conformational space.
\end{itemize}
\par
There lies the strength of the Hamming measure: it is built on meaningful 
structures in higher dimensional space. This is useful for analyzing the 
geometrical features of the clusters obtained: with the help of simple 
statistical data we can extract a lot of information from each one. 
\par
It appears that there are two qualitatively different regions in our 
trajectory. The first one is dominated by an undifferentiated cluster of 
conformations evolving from the initial conditions, ponctuated by incursions 
into substates that may be transient inclusions or appendages.
In the second region the trajectory visits in succesion a series of 
well defined conformational substates.
\par
This analysis seems to indicate that the equilibration time of the system 
may be longer than usually assumed. This same conclusion was reached 
in refs. 1 and 25. We suggest that the information gathered 
be put to use to design better MD simulation protocols.

\newpage
{\bf References}

\begin{itemize}

\item[1.] Gabarro-Arpa, J., Revilla, R. Clustering of a molecular dynamics
          trajectory with a Hamming distance. Comput. and Chem. 24:693-698,
          2000.

\item[2.] Noguti, T. and G$\bar{\textrm{o}}$, N. Structural basis of
          hierarchical multiple substates of a protein. Proteins: Struct.
          Funct. Genet. 5:97,104,113,125,132, 1989.

\item[3.] Frauenfelder, H., Sligar, S.G., Wolynes, P.G. The energy landscapes
          and motions of proteins. Science 254:1598-1603, 1991.

\item[4.] Karpen, M.E., Tobias, D.J. and Brooks III, C.L. Statistical
          clustering techniques for the analysis of long molecular dynamics
          trajectories: analysis of 2.2-ns trajectories of YPGDV. Biochemistry
          32:412-420, 1993.

\item[5.] Shenkin, P.S. and McDonald, Q. Cluster analysis of molecular
          conformations. J. Comput. Chem. 15:889-916, 1994.

\item[6.] Torda, A.E. and van Gunsteren, W.F. Algorithms for clustering
          molecular dynamics configurations. J. of Comput. Chem. 15:1331-1340,
          1994.

\item[7.] Troyer, J.M. and Cohen, F.E. Protein conformational landscapes:
          energy minimization and clustering of a long molecular dynamics
          trajectory. Proteins: Struct. Funct. Genet. 23:97-110, 1995.

\item[8.] Maignan, S., Guilloteau, J. P., Zhou-Liu, Q., Cl\'ement-Mella, C.,
          Mikol, V. Crystal structures of the catalytic domain of HIV-1
          integrase free and complexed with its metal cofactor: high level of
          similarity of the active site with other viral integrases. J. Mol.
          Biol. 282:359-368, 1998.

\item[9.] Pothier, J., Gabarro-Arpa, J., Le Bret, M. MORMIN: a quasi-newtonian
          energy minimizer fitting the nuclear Overhauser data. J. Comp. Chem.
          14:226-236, 1993.

\item[10.] Pearlman, D.A., Case, D.A., Caldwell, J.C., Ross, W.S., 
           Cheatham III, T.E., Ferguson, D.M., Seibel, G.L., Singh, U.C.,
           Weiner, P., Kollman, P.A. AMBER 4.1, University of California,
           San Francisco, 1995.

\item[11.] Cornell, W. D., Cieplak, P., Bayly, C. I., Goulg, I. R., Merz,
           K. M., Ferguson, D. M., Spellmeyer, D. C., Fox, T., Caldwell, J. W.,
           Kollman, P. A. A second generation force field for the simulation 
           of proteins, nucleic acids, and organic molecules. J. Am. Chem. Soc.
           117:5179-5197, 1998.

\item[12.] Aqvist, J. Ion-water interaction potentials derived from free energy
           perturbation simulations. J. Phys. Chem. 94:8021-8024, 1990.

\item[13.] Durden, T.A., York, D., Pedersen, L. Particle mesh Ewald: an
           $N log(N)$ method for Ewald sums in large systems. J. Chem. Phys.
           98:10089, 1993.

\item[14.] Ryckaert, J.P., Cicotti, G., Berendsen, H. Numerical integration
           of the cartesian equations of motion of a system with constaints:
           molecular dynamics of n-alkanes. J. Comput. Phys. 23:327-341, 1977.

\item[15.] Ouali, M., Gousset, H., Geinguenaud, F., Liquier, J., Gabarro-Arpa,
           J., Le Bret, M. and Taillandier, E., Hydration of the
           dT$_n$.dA$_n \times$dT$_n$ parallel triple helix: a Fourier
           transform infrared and gravimetric study correlated with molecular
           dynamics simulations. Nucleic Acids Reseach 25:4816-4824, 1997.

\item[16.] Gabarro-Arpa, J., Le Bret, M., Marcouyoux, A. Cadira, an
           object-oriented platform for modelling molecules and analyzing
           simulations. Comput. and Chem. 21:343-345, 1997.

\item[17.] Becker, O.M. and Karplus, M. The topology of multidimensional
           potential energy surfaces: theory and application to peptide
           structure and kinetics. J. Chem. Phys. 106:1495-1517, 1997.

\item[18.] Nienhaus, G.U., M$\ddot{\textrm{u}}$ller, J.D., McMahon, B.H.,
           Frauenfelder, H., Exploring the conformational energy landscape
           of proteins. Physica D 107:297-311, 1997.

\item[19.] Onuchic, J.N., Luthey-Schulten, Z. and Wolynes, P.G. Theory of
           protein folding: the energy landscape perspective. Annual Review
           of Physical Chemistry 48:545-600, 1997.

\item[20.] Kitao, A., Hayward, S., G$\bar{\textrm{o}}$, N. Energy landscape of
           a native protein: jumping-among-minima model. Proteins: Struct.
           Funct. Genet. 33:496-517, 1998.

\item[21.] Krznaric, D. and Levcopoulos, C. The first subquadratic algorithm 
           for complete linkage clustering. Proc. 6th Annu. Internat. Sympos. 
           Algorithms Comput., Vol. 1004 of Lecture Notes in Computer
           Science, New-York: Springer-Verlag. 1995:392-401.

\item[22.] Kabsch, W. A discussion of the solution for the best rotation to 
           relate two sets of vectors. Acta Cryst. A34:827-828, 1978.

\item[23.] Wlodawer, A., Deisenhofer, J., Huber, R.
           Comparison of two highly refined structures of bovine pancreatic 
           trypsin inhibitor. J. Mol .Biol. 193:145-156, 1987.

\item[24.] Eastman, P., Pellegrini, M., Doniach, S. Protein flexibility in
           solution and in crystals. J. Chem. Phys. 110:10141-10152, 1999.

\item[25.] Ott, E., Chaos in dynamical systems.
           Cambridge University Press, Cambridge (England). 1993.

\end{itemize}

\newpage

\vskip 3mm
\begin{center}
\Large 
\begin{tabular}{| l | l | l | r | l | c |}
\hline
 Cluster   & Start & Length & \multicolumn{1}{c|}{$D$} 
                            & \multicolumn{1}{c|}{$<\Delta rms>$} 
                            & RMSDS \\\hline
 {\bf A}         & 0     & 2.61  &  6.86 &                   &  -  \\
 {\bf A}$_0$     & 0     & 1.4   &  1.94 &                   &  -  \\
 {\bf A}$_0$$_0$ & 0     & 0.045 & 11.07 & 0.027 $\pm$ 0.010 & yes \\
 {\bf A}$_0$$_1$ & 0     & 0.255 &  7.76 & 0.028 $\pm$ 0.012 & yes \\
 {\bf A}$_1$     & 0.665 & 0.135 &  2.06 & 0.031 $\pm$ 0.010 & yes \\
 {\bf A}$_2$     & 0.79  & 0.16  &  0.88 & 0.034 $\pm$ 0.007 & yes \\
 {\bf A}$_3$     & 1.49  & 0.32  &  2.30 & 0.029 $\pm$ 0.011 & yes \\
 {\bf A}$_4$     & 2.015 & 0.43  &  0.94 & 0.025 $\pm$ 0.011 & yes \\
 {\bf A}$_5$     & 2.34  & 0.165 &  1.86 & 0.034 $\pm$ 0.009 & yes \\
 {\bf B}         & 2.61  & 1.395 & 12.83 &                   &  -  \\
 {\bf B}$_1$     & 2.61  & 0.22  &  1.49 & 0.027 $\pm$ 0.012 & yes \\
 {\bf B}$_2$     & 2.745 & 0.165 &  2.05 & 0.030 $\pm$ 0.010 & yes \\
 {\bf B}$_3$     & 2.97  & 0.545 &  0.92 & 0.026 $\pm$ 0.350 & no  \\
 {\bf B}$_4$     & 3.54  & 0.29  &  5.33 & 0.020 $\pm$ 0.009 & yes \\
 {\bf B}$_5$     & 3.83  & 0.175 & 11.67 & 0.028 $\pm$ 0.010 & yes \\\hline

\end{tabular}
\end{center}

\vskip 12mm
\begin{center} \Large{Table I.} \end{center}

\vskip 2mm
In this table we show some parameters for a group of selected clusters 
obtained with the complete-link algoritm and the Hamming distance.
\par
In the {\bf first} column is the name of the cluster. In the present 
nomenclature subclusters inherit the name of the parent cluster, suffixed 
with a number corresponding to their chronological ordering inside the 
parent cluster.
\par
The {\bf second} column is the start of the cluster, in ns from the beginning 
of the simulation. 
The {\bf third} column is the total length (in ns). 
Followed by the specific activity coefficient $D$.
\par
The next two columns refer to the r.m.s. deviation matrix shrinkage test 
(RMSDS) upon minimization: we calculate two r.m.s. deviation matrices for 
the frames belonging to the cluster. The first one, for the original 
structures, the second one, for the minimized structures; in the {\bf fifth} 
column there is the mean value of the difference between the elements 
of the original and the minimized deviation matrices.
\par
The {\bf last} column states the result of the RMSDS test.

\newpage

{\LARGE {\bf Legends of Figures}}

\vskip 12mm
{\bf Figure 1}

\vskip 3mm
Schematic representation of a protein conformational space.
The continuous thin lines are the contour lines of two PESMs: 
{\bf R1} and {\bf R2}.
The straight thick lines represent tuples that cut conformational
space into positive and negative halves.
The broken dotted line represents the protein trajectory, pushing the analogy
a little further one can think of one frame per dot.
The tuples label begins with :
\begin{itemize}
\item[{\bf b }] if they fall approximatively in between the two PESM,
\item[{\bf ns}] if they are non specific, 
\item[{\bf s }] if they are specific,
\item[{\bf sm}] if they cut a PESM approximatively across its center,
\item[{\bf sw}] if they scratch the wall.
\end{itemize}
The approximate sequences associated with the example tuples of the figure are
\begin{itemize}
\item[{\bf b1}]
111111111111111111111111111110010000000000000000000000000000
\item[{\bf b2}]
000000000000000000000000000000000000000100111111111111111111
\item[{\bf ns}]
000001101111111110111111101111001001010000000000000001001101
\item[{\bf sm1}]
100111111111110100000000000000000000000000000000000000000000
\item[{\bf sm2}]
000000000000000000000000000000000000000000000000001110110000
\item[{\bf sw1}]
000000010000000000000000000000000000000000000000000000000000
\item[{\bf sw2}]
000000000000110000000000000000000000000000000000000000000000
\item[{\bf sw3}]
000000000000000000000000000000000000000000000000010000000000
\end{itemize}
The sequences have been scaled so that each bit corresponds approximatively 
to 17 frames (dots) of the trajectory. For each tuple, the 1s are arbitrarily 
assigned to the halfspace where the trajectory spends the less time.

\vskip 12mm
{\bf Figure 2}

\vskip 3mm
\par
Dendogram resulting from the complete-link clustering algorithm with the 
Hamming distance. It has been drawn so that the ordering of the clusters 
along the horizontal axis remains close to the chronological ordering. 
The clusters found with our analysis protocol are underlined. 
The names of subclusters begin with parent cluster's name followed by
a number corresponding to their time ordering within the cluster.
For sake of clarity the diagram has been cut for an arbitrary 
distance since the lower branches do not convey important information.

\newpage
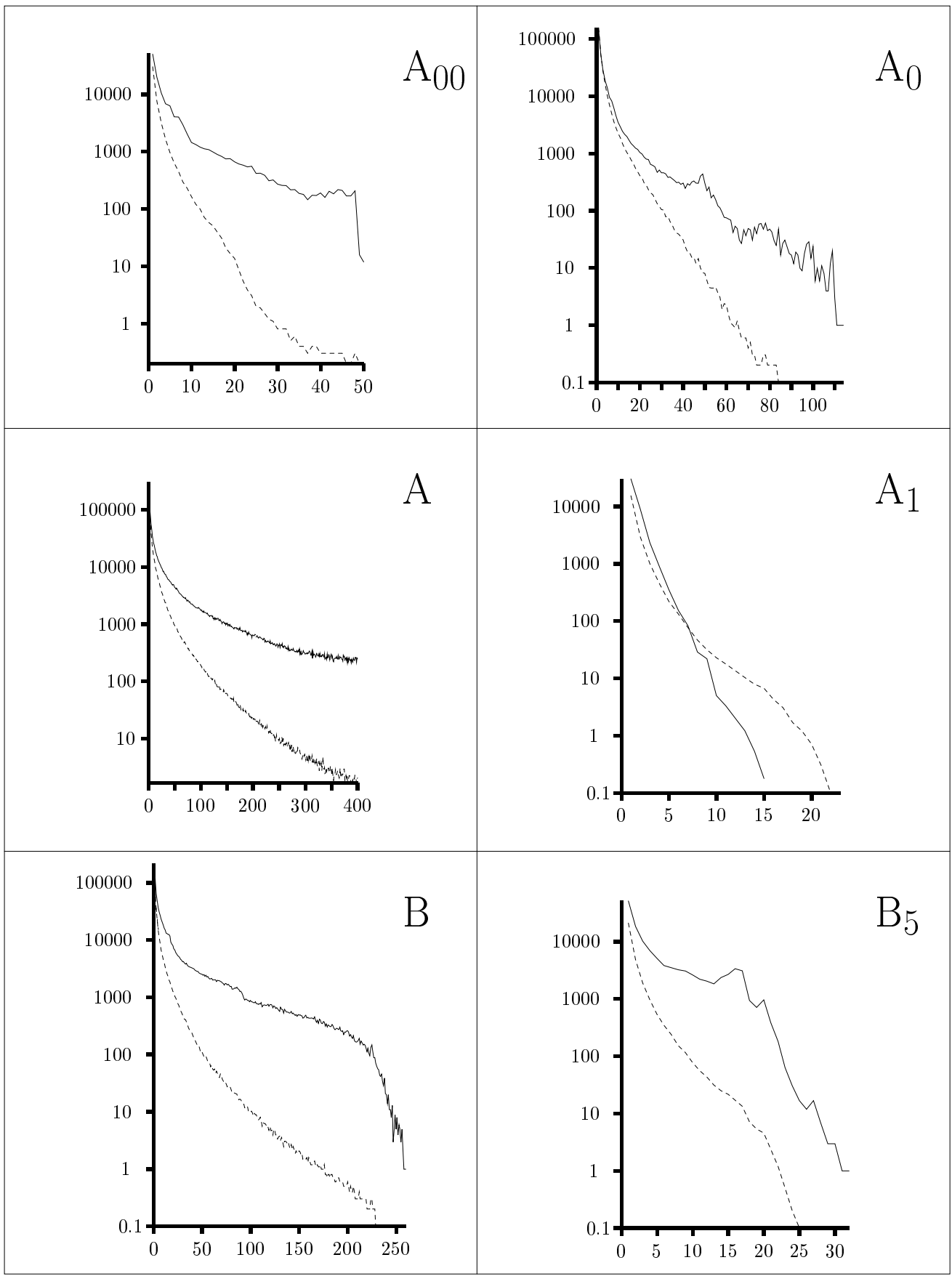
{\bf Figure 3}

\vskip 3mm
Spectrum of specific 4-tuple for some selected clusters.
\par
The vertical coordinate, in logarithmic scale, is the total number of tuples.
The horizontal coordinate is the trajectory total number of frames  
in the tuple minor halfspace. 
\par
For each cluster two curves are plotted :
\begin{itemize}
\item[-] the full line is the spectrum of the specific tuples;
\item[-] the dotted line is the average number spectrum of specific tuples
         on the homologous clusters along the trajectory.
\end{itemize}

\vskip 12mm
{\bf Figure 4}

\vskip 3mm
Matrix density plot.
\par
Lower left  half: r.m.s. deviation matrix.
Upper right half: Hamming distance matrix.
\par
Axis coordinates are all in ns.

\vskip 16mm
{\bf Figure 5}
\vskip 3mm
\par
\begin{itemize}
\item[a)] Dendogram resulting from the complete-link clustering algorithm with 
          the r.m.s. distance. 
\item[b)] Dendogram resulting from the average distance clustering algorithm
          with the r.m.s. distance.
\end{itemize}
For comparison we have underlined the clusters corresponding to clusters
or segments of clusters identified from the dendogram obtained with the Hamming
distance. The symbols have the same meaning as in table I and fig. 2.For sake 
of clarity the diagram has been cut for an arbitrary distance since the lower 
branches do not convey important information.
\par
Clusters bearing a $^\prime$, have more than 80\% homology with their
homologous counterparts in fig. 2.

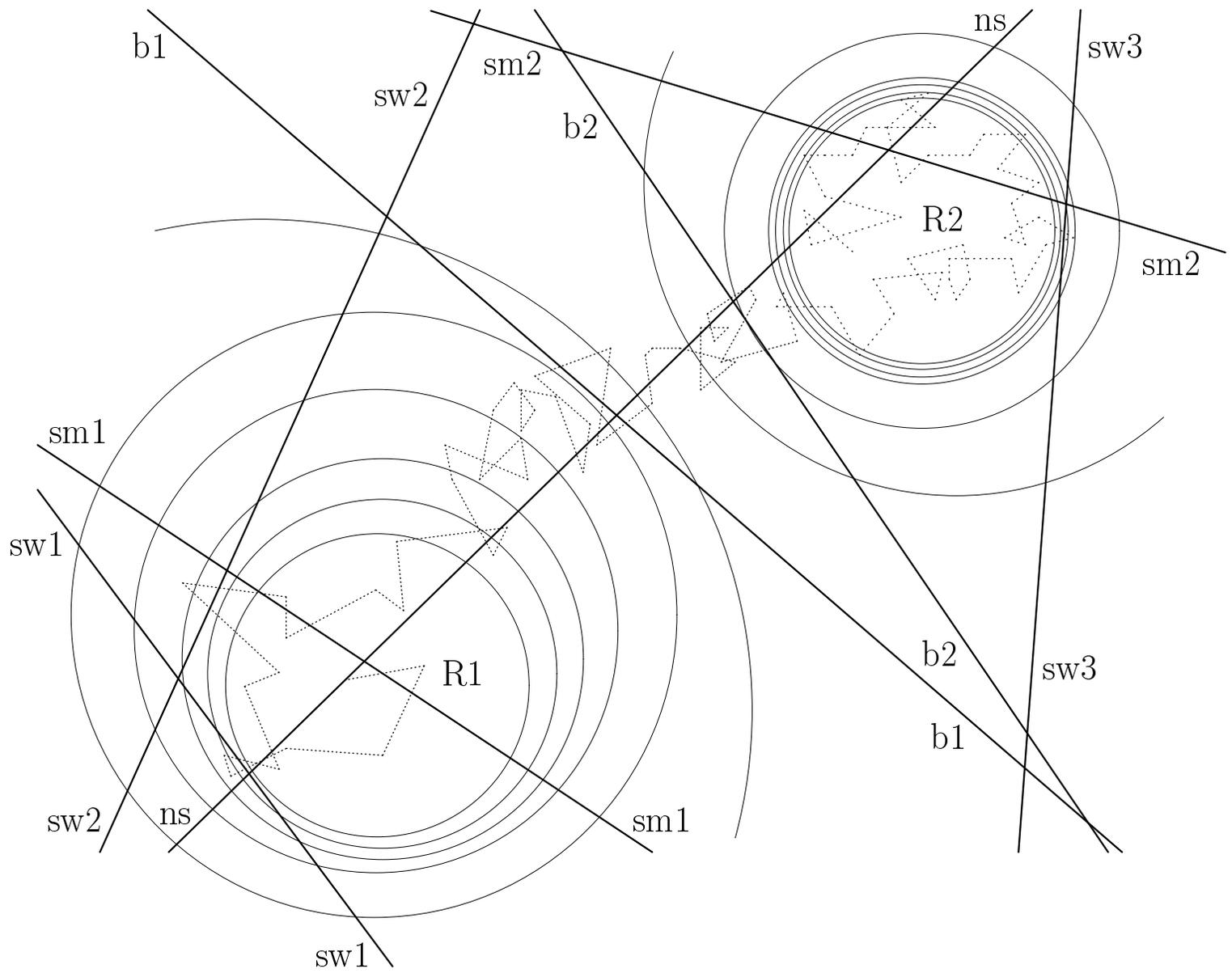
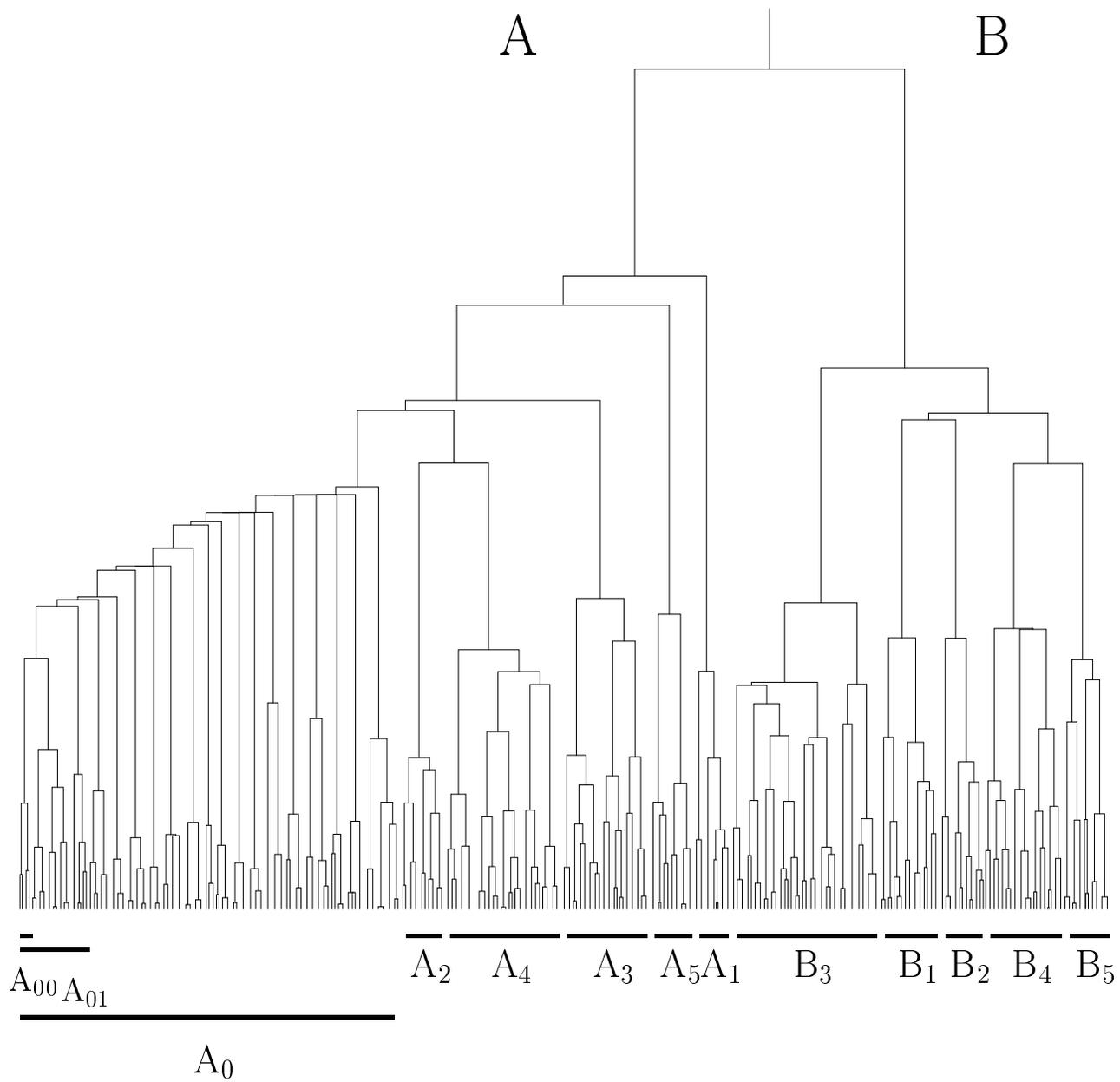
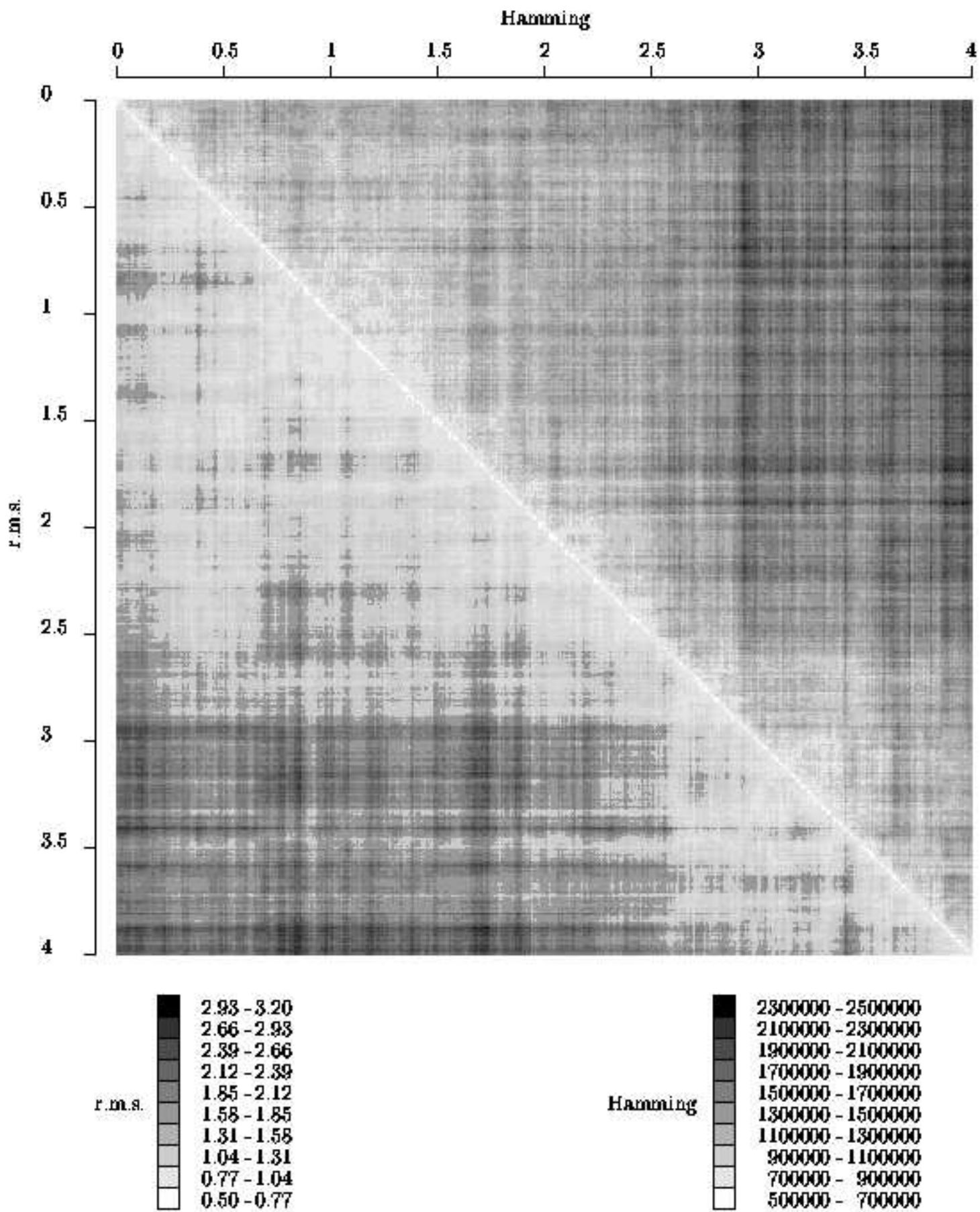
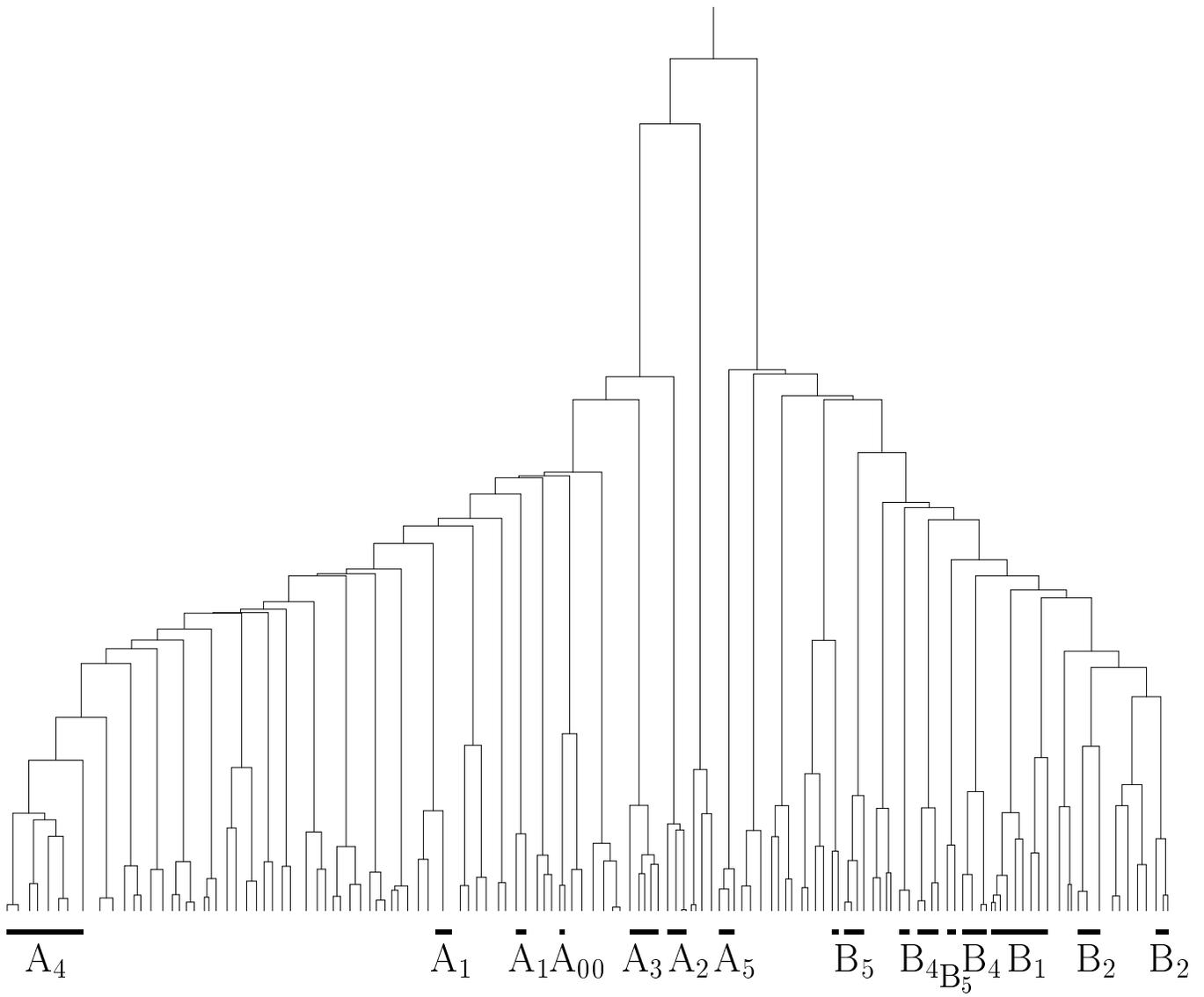
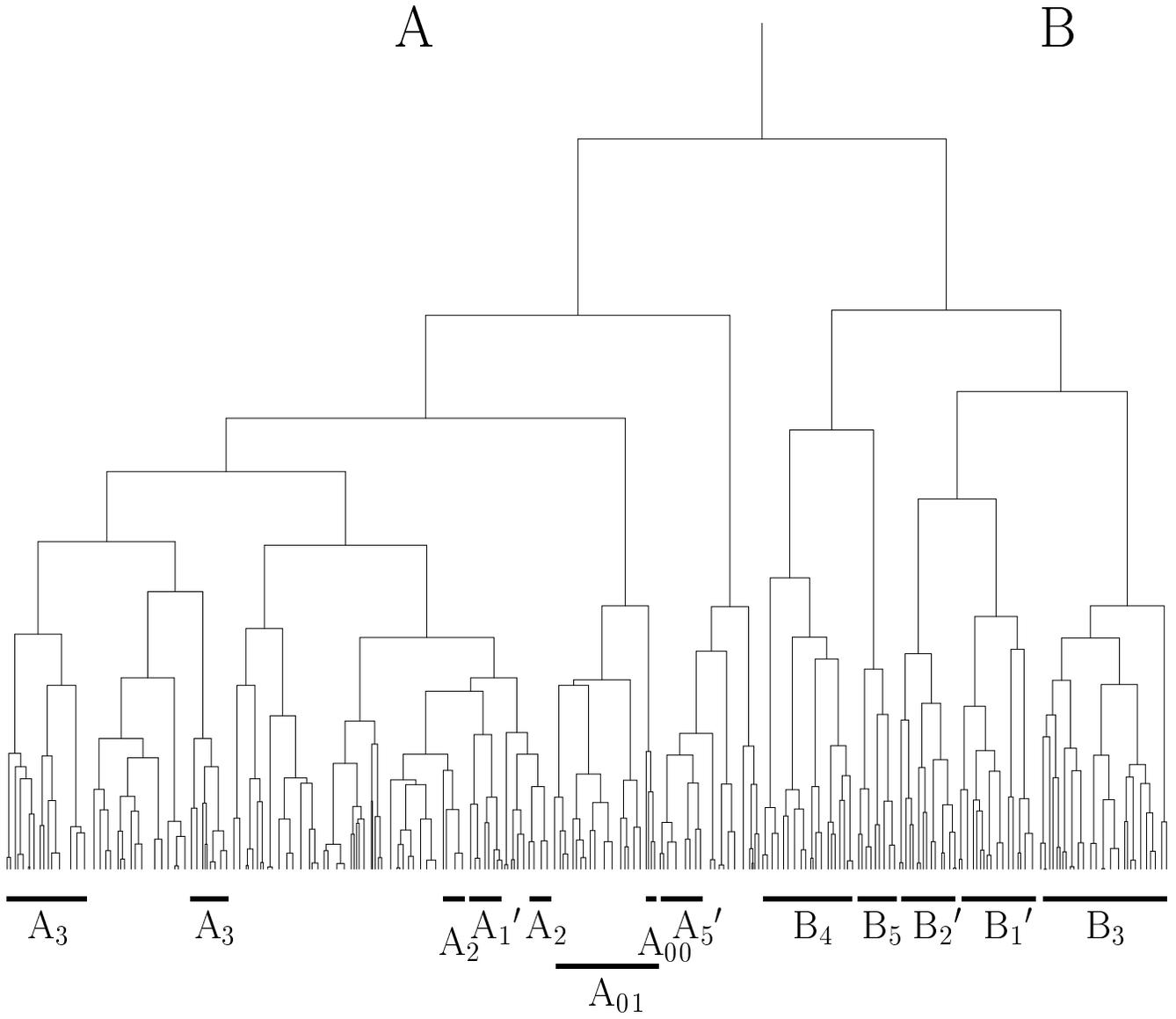
\end{document}